\documentclass[12pt]{article}
\usepackage{graphicx}
\newenvironment{Eqnarray}{\arraycolsep 0.14em\begin{eqnarray}}{\end{eqnarray}}
\def \app{D_{\pi \pi}}

\def \bea{\begin{Eqnarray}}
\def \beq{\begin{equation}}

\def \bl{\bar \lambda}

\def \cn{Collaboration}

\def \eea{\end{Eqnarray}}
\def \eeq{\end{equation}}
\def \ite{{\it et al.}}

\def \s{\sqrt{2}}

\def \3half{\frac{3}{2}}

\begin{document}

\begin{flushright}
TECHNION-PH-2003-37 \\
SLAC-PUB-10178 \\
SCIPP 03/08 \\
EFI 03-43 \\
hep-ph/0310020 \\
\end{flushright}

\renewcommand{\thesection}{\Roman{section}}
\renewcommand{\thetable}{\Roman{table}}
\centerline{\bf Interpreting the time-dependent CP asymmetry in 
$B^0 \to \pi^0 K_S$}
\bigskip

\centerline{Michael Gronau$^{a,b}$, 
Yuval Grossman,$^{a,b,c}$ and
Jonathan L. Rosner$^{d,}$\footnote{On leave from Enrico Fermi Institute
and Department of Physics, University of Chicago, 5640 S. Ellis
Avenue, Chicago, IL 60637}}
\begin{center}
{\it
$^a$Stanford Linear Accelerator Center, \\
  Stanford University, Stanford, CA 94309 \\[3pt] 
$^b$Department of Physics,  
Technion -- Israel Institute of Technology, \\ 
Technion City, 32000
Haifa, Israel \\[3pt]
$^c$Santa Cruz Institute for Particle Physics, \\
University of California, Santa Cruz, CA 95064 \\[3pt]
$^d$Laboratory of Elementary Particle Physics \\
Cornell University, Ithaca, New York 14850
}
\end{center}

\bigskip

\begin{quote}
Flavor SU(3) is used for studying the time-dependent CP asymmetry in 
$B^0 \to \pi^0 K_S$ by relating this process to $B^0\to \pi^0\pi^0$ 
and $B^0 \to K^+K^-$. We calculate correlated bounds on $S_{\pi K} -
\sin 2\beta$ and $C_{\pi K}$, with maximal magnitudes of 0.2 and 0.3,
where $S_{\pi K}$ and $C_{\pi K}$ are coefficients of $\sin\Delta mt$
and $\cos\Delta mt$ in the asymmetry.  Stronger upper limits on $B^0
\to K^+K^-$ are expected to reduce these bounds and to imply nonzero
lower limits on these observables. The asymmetry is studied as a
function of a strong phase and the weak phase $\gamma$.
\end{quote}

\leftline{\qquad PACS codes:  12.15.Hh, 12.15.Ji, 13.25.Hw, 14.40.Nd}

\bigskip
The time-dependent CP asymmetry measured in $B\to J/\psi K_S$ \cite{psiKs}
confirmed the Standard Model, verifying that the Kobayashi-Maskawa phase
\cite{KM} is the dominant origin of CP violation in $K$ and $B$ meson decays.
The theoretical interpretation of this measurement in terms of $\sin 2\beta$,
where $2\beta \equiv {\rm Arg}(V^*_{td})$ is the phase of $B^0 - \bar B^0$
mixing \cite{PDG}, is pure because a single weak phase dominates the weak
amplitude of $B^0 \to J/\psi K_S$ to a high accuracy \cite{Pen}. Charmless
strangeness changing $B^0$ decays into $\phi K_S,~\eta' K_S$ and $K^+K^-K_S$
measured recently \cite{Browder} involve contributions with a second weak phase
which differs from the phase of the dominant penguin amplitude.  This modifies
the time-dependent asymmetries of these processes, which involve hadronic
uncertainties due to the unknown magnitude and strong phase of the small
amplitude relative to the dominant one. Model-independent upper bounds on
these effects were studied using SU(3) or U-spin \cite{GW,GLNQ,GR,CGR,CGLRS}. 
These bounds may be used to indicate when a deviation from the Standard 
Model is observed in asymmetry measurements \cite{newp}.

Recently a first measurement of the CP asymmetry in $B^0(t) \to \pi^0 K_S$ was
reported \cite{Babar},
\beq \label{eqn:SCexp}
S_{\pi K} = 0.48^{+0.38}_{-0.47} \pm 0.11~,~~~~C_{\pi K} = 0.40^{+0.27}_{-0.28}
\pm 0.10~~,
\eeq
where $S_{\pi K}$ and $-C_{\pi K}$ are coefficients of $\sin\Delta mt$ and
$\cos\Delta mt$ terms in the time-dependent asymmetry,
\beq
A(t) \equiv \frac{\Gamma(\bar B^0(t) \to \pi^0 K_S) - \Gamma(B^0(t) \to \pi^0
K_S)}{\Gamma(\bar B^0(t) \to \pi^0 K_S) + \Gamma(B^0(t) \to \pi^0 K_S)}
= -C_{\pi K} \cos(\Delta mt) + S_{\pi K}\sin(\Delta mt)~.
\eeq
The currently measured branching ratio for decays into $\pi^0K^0$, averaged
over $B^0$ and $\bar B^0$, is \cite{FryHFAG}
\beq\label{Bkpi}
\bar{\cal B}(B^0 \to \pi^0 K^0) = (11.92 \pm 1.44)\times 10^{-6}~.
\eeq

In the present Letter we interpret the results for the two asymmetries
$S_{\pi K}$ and $C_{\pi K}$ in terms of the two amplitudes contributing to this
process and their relative strong and weak phases. The relative weak phase
between the two interfering amplitudes is the CKM phase $\gamma \equiv
{\rm Arg} (V^*_{ub})$. Using flavor SU(3), we find a relation
between deviations from $S_{\pi K} = \sin 2\beta$ and $C_{\pi K} = 0$
and decay rates for $B^0\to \pi^0\pi^0$ and $B^0 \to K^+ K^-$.  The
major purpose of this study is to provide, within the CKM 
framework, both upper and lower bounds on these deviations in terms of 
measured rates. It will also be shown how to obtain information about 
$\gamma$ if such deviations are                                       
measured within the range allowed in the Standard Model.

We decompose the amplitude for $B^0\to \pi^0 K^0$ into two terms involving 
CKM factors $V^*_{cb}V_{cs}$ and $V^*_{ub}V_{us}$, which we denote by 
$p'/\s$ and $-c'/\s$, respectively,
\bea\label{pi0K0}
A(B^0 \to \pi^0 K^0)  =
{p' - c'\over \s}~,
\qquad p'  \equiv |p'|e^{i\delta}~, \qquad 
c'\equiv |c'|e^{i\gamma}~.
\eea
This parameterization is true in general within the Standard Model. The
two terms, a penguin amplitude $p'$ with strong phase $\delta$ and a
color-suppressed tree amplitude $c'$ with weak phase $\gamma$, are
graphical representations of SU(3) amplitudes \cite{GHLR} of which we
make use below. The amplitude $p'$ contains color-allowed and
color-suppressed contributions from electroweak penguin operators, $p'
\equiv P' -P'_{\rm EW} - P'^c_{\rm EW}/3$ \cite{EWP}.

Expressions for $S_{\pi K}$ and $C_{\pi K}$ in terms of $p'$ and $c'$ 
can be obtained from definitions, taking into account the negative 
CP eigenvalue of $\pi^0 K_S$ in $B^0$ decays:
\beq
S_{\pi K} \equiv \frac{2{\rm Im}(\lambda_{\pi K})}{1 + |\lambda_{\pi K}|^2}~,
~~~~~~~~~C_{\pi K} \equiv \frac{1 - |\lambda_{\pi K}|^2}
{1 + |\lambda_{\pi K}|^2}~,
\eeq
where
\beq
\lambda_{\pi K} \equiv -e^{-2i\beta}\frac{A(\bar B^0 \to \pi^0 
\bar K^0)}{A(B^0 \to \pi^0 K^0)}~.
\eeq
Using Eq.~(\ref{pi0K0}), the asymmetries 
$S_{\pi K}$ and $C_{\pi K}$ are then written in terms of
$|c'/p'|,~\delta$, $\gamma$, and $\alpha \equiv \pi - \beta - \gamma$, as
\bea
S_{\pi K} & = & {\sin 2\beta - 2|c'/p'| \cos \delta \sin(2 \beta + \gamma) -
|c'/p'|^2 \sin(2 \alpha) \over R_{00}}~ \label{eqn:S},\\[3pt]
C_{\pi K} & = & -{2|c'/p'| \sin \delta \sin \gamma \over R_{00}}~ \label{eqn:C},\\[3pt]
R_{00} & \equiv & 1 - 2|c'/p'| \cos \delta \cos \gamma + |c'/p'|^2~.
\label{eqn:R}
\eea
The amplitudes $p'$ and $c'$ are expected to obey a hierarchy, $|c'| \ll |p'|$
\cite{GHLR,EWP}, which will be justified later on using experimental data.  In
the limit of neglecting $c'$, one has the well-known result $S_{\pi K}
= \sin 2\beta, C_{\pi K}=0$. Keeping only linear terms in $|c'/p'|$,
one has \cite{MG}
\beq\label{SC}
\Delta S_{\pi K} \equiv S_{\pi K} - \sin 2\beta \approx 
-2|c'/p'|\cos 2\beta\cos\delta\sin\gamma~,
~~~~~ C_{\pi K} \approx  - 2|c'/p'|\sin\delta \sin\gamma~.
\eeq
Precise knowledge of the ratio $|c'/p'|$ would permit a determination of
$\sin^2\gamma$ from the two measurements of $S_{\pi
K}$ and $C_{\pi K}$ \cite{GLNQ},
\beq
\sin^2\gamma 
\approx \frac{1}{4|c'/p'|^2}\left(C^2_{\pi K} +
(\Delta S_{\pi K}/\cos 2\beta)^2\right )~.
\eeq

Our goal is to obtain information about $|c'/p'|$ from other $B$ decays 
using flavor SU(3).  For this purpose, we write expressions within flavor
SU(3) for the amplitudes of two strangeness conserving $B^0$ decays
\cite{GHLR,EWP},
\bea\label{pi0pi0}
A(B^0 \to \pi^0\pi^0) & = & (p - c + e + pa)/\s~,\\
\label{K+K-}
A(B^0 \to K^+ K^-) & = & -(e + pa)~~.
\eea
The amplitudes $p$ and $c$ in $\Delta S =0$ decays, defined in analogy
with $p'$ and $c'$ in $\Delta S=1$ decays, involve CKM factors
$V^*_{cb}V_{cd}$ and $V^*_{ub}V_{ud}$, respectively.  The exchange
($e$) and penguin annihilation ($pa$) amplitudes occurring in the
second process are expected to be negligible, unless enhanced by
rescattering \cite{rescatt}.  Current branching ratio measurements,
averaged over $B^0$ and $\bar B^0$, are \cite{FryHFAG}
\bea\label{Bpipi}
\bar{\cal B}(B^0 \to \pi^0\pi^0) & = & (1.89 \pm 0.46)\times 10^{-6}~,\\
\label{Bkk}
\bar{\cal B}(B^0 \to K^+ K^-) & < & 0.6\times 10^{-6}~
(90\%~{\rm confidence~level})~.
\eea
These values already indicate some suppression of $e + pa$ relative to
$p - c$. Using the $90\%$ confidence level upper bound on $\bar{\cal
B}(B^0 \to K^+ K^-)$ and the central value of $\bar{\cal B}(B^0 \to
\pi^0\pi^0)$ we obtain the $90\%$ confidence level bound
\beq\label{bounda}
\frac{|e + pa|^2}{|p-c|^2} \approx \frac{\bar{\cal B}(B^0 \to K^+ K^-)}
{2\bar{\cal B}(B^0 \to \pi^0\pi^0)} \equiv r^2 < 0.16~.
\eeq
Although this suppression is not strong enough to allow neglect of the
terms $e + pa$ in $B^0 \to \pi^0\pi^0$, we will make this
approximation in the majority of our discussion, 
anticipating that the bound (\ref{Bkk}) will be
improved in future measurements of $B^0\to K^+K^-$. 
For completeness, we will also discuss the effect of including the 
amplitude for $B^0 \to K^+K^-$.

The other two terms in $A(B^0 \to \pi^0\pi^0)$, $p$ and $c$, which are
often assumed to dominate this process, are related by SU(3) to the
amplitudes $p'$ and $c'$ in $A(B^0 \to \pi^0 K^0)$ through ratios of
corresponding CKM factors,
\beq\label{pp'cc'}
p = - \bar \lambda\,p'~,~~~~
c = {\bar \lambda}^{-1}\,c'~,~
\eeq
where \cite{PDG}
\beq
\bar \lambda = \frac{V^*_{ub}V_{us}}{V^*_{ub}V_{ud}} =
-\frac{V^*_{cb}V_{cd}}{V^*_{cb}V_{cs}} = \frac{\lambda}{1-\lambda^2/2} =
0.230~.
\eeq
Eqs.~(\ref{pi0pi0}), ({\ref{K+K-}) and (\ref{pp'cc'}) imply
\beq\label{SU3}
A(B^0 \to \pi^0\pi^0) + A(B^0 \to K^+ K^-)/\s =
(- \bar \lambda\,p' - {\bar \lambda}^{-1}\,c')/\s~.
\eeq
This relation between $A(B^0 \to \pi^0\pi^0)$, $A(B^0 \to K^+
K^-)$ and $A(B^0\to \pi^0 K^0)$ in
(\ref{pi0K0}), which involves the same hadronic amplitudes $p'$ and
$c'$ with different CKM coefficients, is the basis of our study. 
We emphasize that it follows purely from SU(3), as can be read form the
tables in \cite{GLNQ,GHLR}.

We start by neglecting the $B^0 \to K^+ K^-$ amplitude.
Under this approximation, using Eqs.~(\ref{pi0K0}) and (\ref{SU3}), we
calculate the ratio of rates for decays into $\pi^0\pi^0$ and $\pi^0
K^0$, averaged over $B^0$ and $\bar B^0$ and multiplied by $\bl^2$,
\beq\label{Rpi/K}
R_{\pi/K} \equiv
\frac{\bl^2\bar{\cal B}(B^0 \to \pi^0\pi^0)}{\bar{\cal B}(B^0 \to \pi^0K^0)}
= \frac{|c'/p'|^2 + \bl^4 + 2\bl^2|c'/p'|\cos\delta\cos\gamma}
{1 + |c'/p'|^2 - 2|c'/p'|\cos\delta\cos\gamma}~.
\eeq
The current experimental value of the ratio $R_{\pi/K}$ obtained from
(\ref{Bkpi}) and (\ref{Bpipi}) is
\beq \label{eqn:Rvalue}
R_{\pi/K} = 0.0084 \pm 0.0023~.
\eeq
For a given value of $R_{\pi/K}$ in this range, $|c'/p'|$ is a monotonically
decreasing function of $\cos\delta\cos\gamma$,
\beq\label{c'/p'}
|c'/p'| ={\sqrt{[(\bl^2 + R_{\pi/K})\cos\delta\cos\gamma]^2 +
(1-R_{\pi/K})(R_{\pi/K} - \bl^4)} - (\bl^2 +
R_{\pi/K})\cos\delta\cos\gamma
\over (1-R_{\pi/K})}
.
\eeq
Eq.~(\ref{Rpi/K}) can be used to set bounds on $|c'/p'|$. Noting that
$-1 \le \cos\delta\cos\gamma \le 1$,  one has
\beq
\left (\frac{|c'/p'| - \bl^2}{1 + |c'/p'|}\right )^2 \le R_{\pi/K} \le
\left (\frac{|c'/p'| + \bl^2}{1 - |c'/p'|}\right )^2~.
\eeq
With $\sqrt{R_{\pi/K}} = 0.091 \pm 0.012$, one finds
\beq\label{boundsc'/p'}
0.035 \pm 0.011 = \frac{\sqrt{R_{\pi/K}} - \bl^2}{1 + \sqrt{R_{\pi/K}}}
\le |c'/p'| \le \frac{\sqrt{R_{\pi/K}} + \bl^2}{1 - \sqrt{R_{\pi/K}}}
 = 0.158 \pm 0.016~.
\eeq
This implies the following bounds at $95\%$ confidence level:
\beq\label{c'/p'bounds}
 0.02 \le |c'/p'| \le 0.18~.  \eeq The lower and upper bounds
 correspond to $\cos\delta\cos\gamma = 1$ and $\cos\delta\cos\gamma =
 -1$, respectively. Slightly stronger bounds on $|c'/p'|$ may be
 obtained by using current constraints on CKM parameters
 \cite{CKMfitter} implying $\gamma > 38^\circ$, or $-0.79 \le
 \cos\delta\cos\gamma \le 0.79$, at 95$\%$ confidence level.

We now turn to $\Delta S_{\pi K}$ and $C_{\pi K}$ 
for which we wish to calculate bounds. We proceed in two ways. First, 
we use the approximate expressions (\ref{SC}) and derive analytically
separate bounds on these two measurables. Then we use the exact 
expressions (\ref{eqn:S})--(\ref{eqn:R}) in order to draw a graphical 
plot for correlated bounds.

Eqs.~(\ref{SC}) and (\ref{c'/p'}) may be used to calculate maxima for the 
magnitudes of $\Delta S_{\pi K}$ and $C_{\pi K}$ when varying $\delta$ 
and $\gamma$ for
fixed values of $\beta$ and $R_{\pi/K}$.  Since $|c'/p'|$ decreases
monotonically with $\cos\delta\cos\gamma$, the maximum of $\Delta S_{\pi K}$
which is proportional to $\cos\delta$ is obtained for
$\delta = \pi$ and is positive. 
As for $\gamma$, the maximum is obtained for a value given approximately by
\beq \label{gammax}
\tan\gamma \simeq \frac{\sqrt{R_{\pi/K}}}{\bl^2 + R_{\pi/K}}~.
\eeq
The current data imply a value $\gamma \approx 56^\circ$, which lies in 
the allowed range \cite{CKMfitter} $38^\circ < \gamma < 80^\circ$.
Using the central values, $\beta = 23.7^\circ$ \cite {CKMfitter} and
$R_{\pi/K} = 0.0084$, the following maximal positive value is obtained
for $\Delta S_{\pi K}$:
\beq\label{Smax}
[\Delta S_{\pi K}]_{\rm max} \approx 0.13~.
\eeq
The most negative value of this measurable in the allowed region of 
$\gamma$ is obtained for $\delta =0$ and $\gamma = 80^\circ$,
\beq\label{Smin}
[\Delta S_{\pi K}]_{\rm min} \approx -0.09~.
\eeq 

Since $C_{\pi K}(-\delta) = -C_{\pi K}(\delta)$, one may consider only 
its magnitude. The maximum of $|C_{\pi K}|$ is obtained at 
$\delta =\gamma= \pi/2$, for which one finds
\beq\label{Cmax}
|C_{\pi K}|_{\rm max} \approx 2\sqrt{R_{\pi/K} - \bl^4} = 0.15~.
\eeq
The value of $|C_{\pi K}|_{\rm max}$ is essentially the same at
$\gamma = 80^\circ$.  We will comment on this maximal value below,
where we relate it to the CP asymmetry in $B^0 \to \pi^0\pi^0$.

The exact expressions (\ref{eqn:S})--(\ref{eqn:R}) 
imply correlated constraints in the
$S_{\pi K}$--$|C_{\pi K}|$ plane associated with fixed values of $R_{\pi/K}$.  
We take values of $\delta$ with a $15^\circ$ step, values of $\gamma$ 
satisfying \cite{CKMfitter} $38^\circ \le \gamma \le 80^\circ$, and values 
of $R_{\pi/K}$ between the $\pm 1 \sigma$ limits of Eq.\ (\ref{eqn:Rvalue}).  
A scatter plot of the results is shown in Fig.\ \ref{fig:scan}.  
We find
\beq \label{eqn:scan}
-0.11 \le \Delta S_{\pi K} \le 0.12~~,~~~|C_{\pi K}| \le 0.17~~~.
\eeq
The bounds of the allowed region differ only slightly from 
(\ref{Smax})--(\ref{Cmax}), for which approximate expressions were 
used and a central value was chosen for $R_{\pi/K}$.
An important point demonstrated by the plot is that the measurement of 
$B^0 \to \pi^0 \pi^0$ is seen to imply a {\it minimum} deviation from the 
point $(S_{\pi K},C_{\pi K}) = (\sin2 \beta,0)$, 
which requires a non-zero value for $|c'/p'|$.

SU(3) breaking in the ratios $p'/p$ and $c'/c$ is expected to introduce
corrections at a level of 20--30 $\%$ in these ratios. These effects may be
studied using QCD calculations \cite{BBNS,KLS}. Corresponding effects in
$\Delta S_{\pi K}$ and $|C_{\pi K}|$ are likely to be smaller, since
these two quantities involve the ratio of amplitudes $|c'/p'|$ in which 
some SU(3) breaking corrections are expected to cancel.  We conclude that 
$|\Delta S_{\pi K}|$ and $|C_{\pi K}|$ are at most as large as 0.2. Larger
values would signal physics beyond the Standard Model in $B^0 \to
\pi^0 K^0$.  The possible role of new physics in
$B\to \pi K$ decays was studied in \cite{KpiAnomaly}.

\null
\begin{figure}[t]
\includegraphics[height=4.8in]{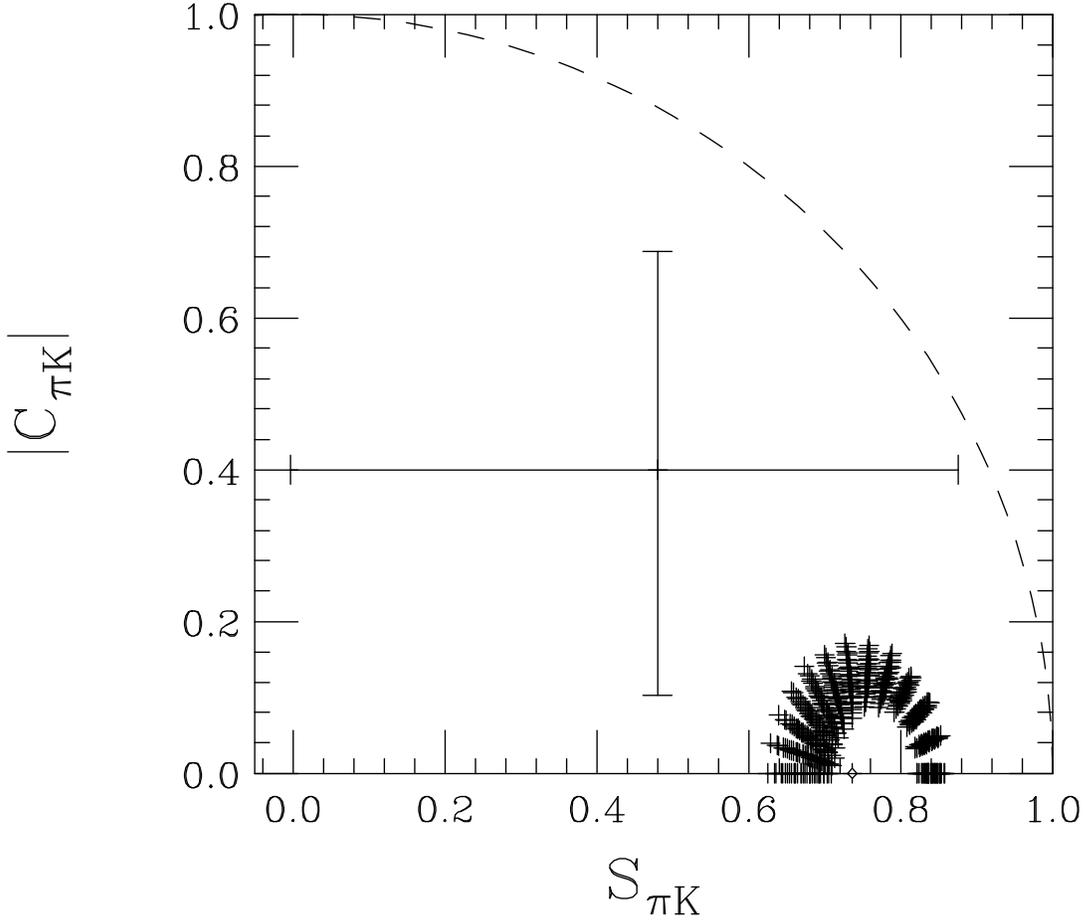}
\caption{Points in the $S_{\pi K}$--$|C_{\pi K}|$ plane
satisfying $\pm 1 \sigma$ limits (\ref{eqn:Rvalue}) on the ratio $R_{\pi/K}$.
The small plotted point denotes the pure-penguin value $S_{\pi K} =
\sin2 \beta$, $C_{\pi K} = 0$.  The point with large error bars denotes
the experimental value (\ref{eqn:SCexp}).  The dashed arc denotes the
boundary of allowed values:  $S_{\pi K}^2 + C_{\pi K}^2 \le 1$.  (Lowest,
highest) values of $|\delta|$ correspond to (lowest, highest) values of
$S_{\pi K}$.  (Lowest, highest) values of $\gamma$ correspond in general to
(innermost, outermost) ellipses.
\label{fig:scan}}
\end{figure}
 
Note that the maximal values of $|\Delta S_{\pi K}|$ and
$|C_{\pi K}|$ are obtained for different values of $\delta$.
Measuring nonzero values for $\Delta S_{\pi K}$ and 
$C_{\pi K}$, within the above bounds permitted by the Standard Model,
could be used to obtain information about $\tan\delta$ and 
$|c'/p'|\sin\gamma$ through rather simple expressions
obtained in the linear approximation (\ref{SC}),
\beq
\tan\delta \approx \frac{C_{\pi K}\cos 2\beta}{\Delta S_{\pi K}}~,
~~~~|c'/p'|\sin\gamma \approx - \frac{C_{\pi K}}{2\sin\delta}~.
\eeq
Since $|c'/p'|$ in (\ref{c'/p'}) depends on $\cos\delta\cos\gamma$,
this can in principle be used to determine $\gamma$ up to discrete
ambiguities.

In the above calculation we neglected the contribution of $A(B^0 \to 
K^+K^-)$ to the left-hand-side of Eq.~(\ref{SU3}), anticipating 
that the upper bound on the corresponding branching ratio (\ref{Bkk}) 
will be improved in the future. Including this contribution introduces 
several unknowns related to magnitudes and strong phases of the terms
$e$ and $pa$, 
but nevertheless permits a similar analysis of correlated bounds on   
the asymmetries $\Delta S_{\pi K}$ and $C_{\pi K}$ in terms of the 
strong phase $\delta$ between $p'$ and $c'$ and the weak phase $\gamma$. 
That is, one may compute the maximal allowed values of $|c'/p'|$, 
$|\Delta S_{\pi K}|$ and $|C_{\pi K}|$ as functions of $\delta$ and 
$\gamma$ under the current bound (\ref{Bkk}). 

Starting from Eq.~(\ref{SU3}), one forms a ratio 
\beq
R'_{\pi/K} \equiv \frac{\bl^2\,[|A_{\pi\pi} + 
A_{KK}/\s|^2 + |\bar A_{\pi\pi} + 
\bar A_{KK}/\s|^2]}{|A_{\pi K}|^2 + |\bar A_{\pi K}|^2}~~,
\eeq
where $A_{\pi\pi,KK,\pi K} \equiv A(B^0 \to 
\pi^0\pi^0,K^+K^-,\pi^0 K^0)$ and $\bar A_{\pi\pi,KK,\pi K}$ are 
the amplitudes of the charge-conjugate processes. 
This ratio is given by the right-hand-side of 
Eq.~(\ref{Rpi/K}) in terms of $|c'/p'|,~\delta$ and $\gamma$. 
The maximal and minimal allowed values of $|c'/p'|$ 
are attained for the largest and smallest possible values of 
$R'_{\pi/K}$, respectively, and are calculated from expressions 
similar to Eq.~(\ref{boundsc'/p'}), in which values of $R_{\pi/K}$ 
are replaced by corresponding values of $R'_{\pi/K}$. The maximal 
values of $|\Delta S_{\pi K}|$ and $|C_{\pi K}|$ correspond to the 
maximum of $R'_{\pi/K}$.

Although $R'_{\pi/K}$ is not measurable, upper and lower bounds on
this quantity follow from the general inequalities
\bea
& & 
\left (\sqrt{|A_{\pi\pi}|^2 + |\bar A_{\pi\pi}|^2} -
\sqrt{(|A_{KK}|^2 + |\bar A_{KK}|^2)/2}\right)^2\\
& &  
\le
|A_{\pi\pi} + A_{KK}/\s|^2 + |\bar A_{\pi\pi} + \bar A_{KK}/\s|^2\\
& & 
\le 
\left (\sqrt{|A_{\pi\pi}|^2 + |\bar A_{\pi\pi}|^2} +
\sqrt{(|A_{KK}|^2 + |\bar A_{KK}|^2)/2}\right)^2~.
\eea
The left and right side inequalities become equalities when 
$A_{KK}/\s = \mp r A_{\pi\pi}$ and $\bar A_{KK}/\s = 
\mp r \bar A_{\pi\pi}$, where $r$ is defined in Eq.\ (\ref{bounda}). 
Denoting
\beq
R'_{\pm} \equiv
\bl^2\left(\sqrt{\frac{\bar{\cal B}(B^0 \to \pi^0\pi^0)}
{\bar{\cal B}(B^0 \to \pi^0K^0)}} \pm
\sqrt{\frac{\bar{\cal B}(B^0 \to K^+K^-)}
{2\bar{\cal B}(B^0 \to \pi^0K^0)}}\right)^2
= R_{\pi/K} (1 \pm r)^2~,
\eeq
one then has
\beq
R'_- \le R'_{\pi/K} \le R'_+~.
\eeq
Thus, we can use the measured limits on $R'_{\pm}$ to set bounds on 
$\Delta S_{\pi K}$ and $C_{\pi K}$ in the same way
as before, with $B^0 \to K^+ K^-$ now taken into account.  We replace the
upper bound on $R_{\pi/K}$ by $R'_+ = (1+r_{\rm max})^2 R_{\pi/K}$, 
and the lower bound by $R'_- = (1-r_{\rm max})^2 R_{\pi/K}$, 
where $r_{\rm max} = 0.4$ from Eq.\ (\ref{bounda}).

Using the central values of the measured rates of $\bar{\cal B}(B^0 \to
\pi^0\pi^0)$ and $\bar{\cal B}(B^0 \to \pi^0K^0)$ and the upper bound on
$\bar{\cal B}(B^0 \to K^+K^-)$ we get
\beq
R'_+ = 0.016~,~~~~~~~R'_- = 0.003~. 
\eeq 
An equation similar to (\ref{boundsc'/p'}), in which $R_{\pi/K}$ is 
replaced by $R'_+$ for an 
upper bound on $|c'/p'|$, and by $R'_-$ 
for a lower bound, implies
\beq
0.002 \le |c'/p'| \le 0.21~.
\eeq
Including errors in $R_{\pi/K}$ allows a value $|c'/p'|=0$, implying
that $\Delta S_{\pi K} = C_{\pi K} = 0$ is not forbidden in contrast 
to the case of neglecting the amplitude for $B^0 \to K^+K^-$.

The above value of $R'_+$ implies, for $\delta = \pi$ 
and $\gamma \simeq 61^\circ$ given by (\ref{gammax}) (in which $R_{\pi/K}$
is replaced by $R'_+$),
\beq
[\Delta S_{\pi K}]_{\rm max} \approx 0.19~,
\eeq
while for $\delta = 0$ and $\gamma = 80^\circ$ we find
\beq
[\Delta S_{\pi K}]_{\rm min} \approx -0.14~.
\eeq
We also obtain 
\beq
|C_{\pi K}|_{\rm max} \approx 2\sqrt{R'_+ - \bl^4} = 0.23~.
\eeq

The allowed range of $S_{\pi K}$ and $C_{\pi K}$ can be calculated 
using the exact expressions (\ref{eqn:S})--(\ref{eqn:R}), taking
account of the possible contribution of $B^0 \to K^+ K^-$.  One replaces the
range $6.1 \le (R_{\pi K}/10^{-3}) \le 10.7$ by $2.2 \le (R'_{\pi K}/10^{-3})
\le 20.9$, where $2.2 = (1-r_{\rm max})^2 6.1$ and $20.9 = (1+r_{\rm max})^2
10.7$.  The result is shown in Fig.\ \ref{fig:bigscan}.  
The bounds 
(\ref{eqn:scan}) are replaced by
\beq \label{eqn:bigscan}
-0.18 \le \Delta S_{\pi K} \le 0.16~~,~~~|C_{\pi K}| \le 0.26~,
\eeq
where extreme values are larger than those in (\ref{eqn:scan}) by about 
$50\%$. As mentioned, there is now no minimum deviation from the point 
$(S_{\pi K},C_{\pi K}) = (\sin 2 \beta, 0)$. Such a deviation is  
expected when improving the upper bound on $B^0 \to K^+K^-$. 

We wish to conclude with a few comments:

\begin{itemize}
\item In the first part of our study we have neglected $A(B^0\to K^+K^-)/\s$ 
relative to $A(B^0\to \pi^0\pi^0)$.  As we have shown now, including 
the first amplitude weakens somewhat the upper bounds on $|c'/p'|$ and on
$|\Delta S_{\pi K}|$ and $|C_{\pi K}|$. We expect that in the next few years
the current bound (\ref{bounda}) will be improved to imply $|e + pa|/|p-c| <
0.2-0.3$.  At this point, the approximation of neglecting these terms
will introduce an uncertainty at the same level as SU(3) breaking
corrections in $p'/p$ and $c'/c$.  It would be interesting to study the
magnitude of $e + pa$ and SU(3) breaking effects in the above ratios by using
QCD calculations \cite{BBNS,KLS}.
\item We considered only the direct CP asymmetry $-C_{\pi K}$ in 
$B^0\to \pi^0 K^0$.  Eventually, one hopes to also measure an
asymmetry in $B^0 \to \pi^0\pi^0$.  In the SU(3) approximation and
neglecting $e + pa$, the CP rate differences in these two processes
have equal magnitudes and opposite signs \cite{Uspin}.  Measuring the
two asymmetries may be used to check for SU(3) breaking corrections.
Since the charge averaged rate of $B^0\to \pi^0 K^0$ is about six
times larger than that of $B^0 \to \pi^0\pi^0$, a small asymmetry
$C_{\pi K}$ implies a six times larger asymmetry in decays to
$\pi^0\pi^0$. The maximal value calculated for $C_{\pi K}$ in
(\ref{Cmax}) corresponds to an asymmetry of about 100$\%$ in $B^0 \to
\pi^0\pi^0$. Turning things around, an absolute maximal 100$\%$
asymmetry in $B^0\to \pi^0\pi^0$ implies in the SU(3) limit a maximal
asymmetry of 0.15 in $B^0\to \pi^0 K^0$ as calculated in (\ref{Cmax}).

\null
\begin{figure}[t]
\includegraphics[height=5.1in]{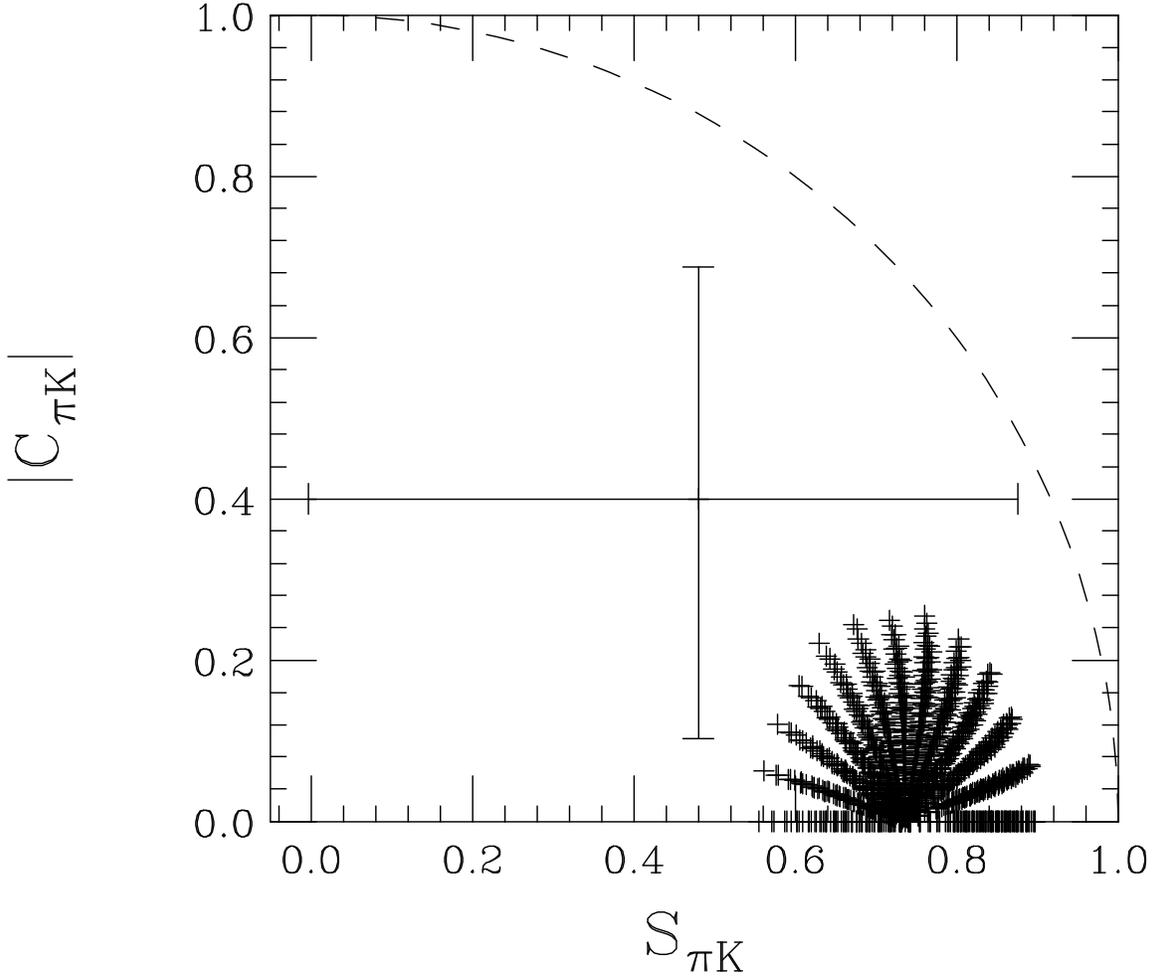}
\caption{Points in the $S_{\pi K}$--$|C_{\pi K}|$ plane satisfying $\pm 1
\sigma$ limits on the ratio $(1\pm r_{\rm max})^2R_{\pi/K}$,
where $r_{\rm max} = 0.4$, i.e., taking into account upper bound on
$\bar{\cal B}(B^0 \to K^+ K^-)$.  Other notation is the same as in
Fig.\ \ref{fig:scan}.
\label{fig:bigscan}}
\end{figure}

\item The process $B^0 \to \pi^0 K^0$ is related by 
U-spin to $B_s \to
\pi^0\bar K^0$ \cite{Uspin}, for which the amplitude is given by 
\cite{GHLR}
\beq
A(B_s \to \pi^0 \bar K^0) = (p - c )/\s~.
\eeq
In the SU(3) limit, this amplitude is equal to $A(B^0 \to \pi^0\pi^0) +
A(B^0 \to K^+ K^-)/\s$ and may replace this sum on the left-hand-side 
of Eq.~(\ref{SU3}). In order to obtain bounds on $S_{\pi K}$ and 
$C_{\pi K}$ as above, one would then have to know 
the ratio $\bar {\cal B}(B_s \to \pi^0 \bar K^0)/\bar {\cal B}(B^0 \to 
\pi^0 K^0)$.
Measuring the charge averaged rate for $B_s \to \pi^0 \bar K^0$ 
in an environment of a hadronic collider may be quite challenging.
\item
The method for obtaining correlated bounds on $\Delta S_{\pi K}$ and
$C_{\pi K}$ may be applied to CP asymmetries in other processes, such as
$B^0 \to \eta' K_S$ and $B^0 \to \phi K_S$.  In \cite{GLNQ} upper bounds
on quantities analogous to $|c'/p'|$ were obtained by relating 
within SU(3) the amplitudes of these processes to the sum of several $\Delta
S=0$ amplitudes. 
For $B^0 \to \phi K_S$, the bound requires an assumption that a term with 
weak phase $\gamma$ is not much larger than in $B^+ \to \phi K^+$.
The SU(3) relations for $B^0 \to \eta' K_S$ and $B^+ \to \phi K^+$ were 
shown to follow from U-spin symmetry \cite{CGR,CGLRS}. The bounds on a ratio
analogous to $|c'/p'|$ provided estimates for 
the maximal values of the asymmetries $|S - \sin 2\beta|$.
In deriving these bounds additive corrections of order $(\lambda^2)$ were 
neglected in quantities resembling $\sqrt{R_{\pi/K}}$, and only leading 
order terms in a $|c'/p'|$ expansion were kept. Studying the dependence 
of the asymmetries $S$ and $C$ on $c'/p'$, and on strong and weak phases, 
and avoiding such approximations, one can use the SU(3) relations of 
\cite{GLNQ,CGR,CGLRS} in order to get more precise bounds in the $S-|C|$ plane.
\end{itemize}

\bigskip
We are grateful to Jim Smith for raising a question which motivated
this study.  M. G. wishes to thank the SLAC theory group for its kind
hospitality. The work of Y. G. is supported in part by a grant from 
GIF, the German--Israeli Foundation for Scientific Research and
Development, by the United States--Israel Binational Science
Foundation through grant No.~2000133, by the Israel Science Foundation
under grant No.~237/01, by the Department of Energy, contract
DE-AC03-76SF00515 and by the Department of Energy under grant
no.~DE-FG03-92ER40689.

\def \ajp#1#2#3{Am.\ J. Phys.\ {\bf#1}, #2 (#3)}
\def \apny#1#2#3{Ann.\ Phys.\ (N.Y.) {\bf#1}, #2 (#3)}
\def \app#1#2#3{Acta Phys.\ Polonica {\bf#1}, #2 (#3)}
\def \arnps#1#2#3{Ann.\ Rev.\ Nucl.\ Part.\ Sci.\ {\bf#1}, #2 (#3)}
\def \art{and references therein}
\def \cmts#1#2#3{Comments on Nucl.\ Part.\ Phys.\ {\bf#1}, #2 (#3)}
\def \cn{Collaboration}
\def \cp89{{\it CP Violation,} edited by C. Jarlskog (World Scientific,
Singapore, 1989)}
\def \econf#1#2#3{Electronic Conference Proceedings {\bf#1}, #2 (#3)}
\def \efi{Enrico Fermi Institute Report No.}
\def \epjc#1#2#3{Eur.\ Phys.\ J.\ C {\bf#1}, #2 (#3)}
\def \ib{{\it ibid.}~}
\def \ibj#1#2#3{~{\bf#1}, #2 (#3)}
\def \ijmpa#1#2#3{Int.\ J.\ Mod.\ Phys.\ A {\bf#1}, #2 (#3)}
\def \ite{{\it et al.}}
\def \jhep#1#2#3{JHEP {\bf#1}, #2 (#3)}
\def \jpb#1#2#3{J.\ Phys.\ B {\bf#1}, #2 (#3)}
\def \mpla#1#2#3{Mod.\ Phys.\ Lett.\ A {\bf#1} (#3) #2}
\def \nat#1#2#3{Nature {\bf#1}, #2 (#3)}
\def \nc#1#2#3{Nuovo Cim.\ {\bf#1}, #2 (#3)}
\def \nima#1#2#3{Nucl.\ Instr.\ Meth.\ A {\bf#1}, #2 (#3)}
\def \npb#1#2#3{Nucl.\ Phys.\ B~{\bf#1}, #2 (#3)}
\def \npps#1#2#3{Nucl.\ Phys.\ Proc.\ Suppl.\ {\bf#1}, #2 (#3)}
\def \PDG{Particle Data Group, K. Hagiwara \ite, 
\prd{66}{010001}{2002}}
\def \pisma#1#2#3#4{Pis'ma Zh.\ Eksp.\ Teor.\ Fiz.\ {\bf#1}, 
#2 (#3) [JETP
Lett.\ {\bf#1}, #4 (#3)]}
\def \pl#1#2#3{Phys.\ Lett.\ {\bf#1}, #2 (#3)}
\def \pla#1#2#3{Phys.\ Lett.\ A {\bf#1}, #2 (#3)}
\def \plb#1#2#3{Phys.\ Lett.\ B {\bf#1}, #2 (#3)}
\def \prl#1#2#3{Phys.\ Rev.\ Lett.\ {\bf#1}, #2 (#3)}
\def \prd#1#2#3{Phys.\ Rev.\ D\ {\bf#1}, #2 (#3)}
\def \prp#1#2#3{Phys.\ Rep.\ {\bf#1}, #2 (#3)}
\def \ptp#1#2#3{Prog.\ Theor.\ Phys.\ {\bf#1}, #2 (#3)}
\def \rmp#1#2#3{Rev.\ Mod.\ Phys.\ {\bf#1}, #2 (#3)}
\def \yaf#1#2#3#4{Yad.\ Fiz.\ {\bf#1}, #2 (#3) [Sov.\ 
J.\ Nucl.\ Phys.\
{\bf #1}, #4 (#3)]}
\def \zhetf#1#2#3#4#5#6{Zh.\ Eksp.\ Teor.\ Fiz.\ {\bf #1}, 
#2 (#3) [Sov.\
Phys.\ - JETP {\bf #4}, #5 (#6)]}
\def \zpc#1#2#3{Zeit.\ Phys.\ C {\bf#1}, #2 (#3)}
\def \zpd#1#2#3{Zeit.\ Phys.\ D {\bf#1}, #2 (#3)}

\end{document}